\documentclass[oldversion]{aa} 
\usepackage{graphicx}
\usepackage{txfonts}

\usepackage{natbib}
\bibpunct{(}{)}{;}{a}{}{,} 

\begin{document}
   \title{X-ray power law spectra in active galactic nuclei}

   \subtitle{}

   \author{W. Ishibashi
          \inst{1,2}
          \and
          T. J.-L. Courvoisier
          \inst{1,2}
          }

   \institute{ ISDC Data Centre for Astrophysics, ch. d'Ecogia 16, 1290 Versoix, Switzerland 
    \and Geneva Observatory, Geneva University, ch. des Maillettes 51, 1290 Sauverny, Switzerland \\
    e-mail: Wakiko.Ishibashi@unige.ch, Thierry.Courvoisier@unige.ch 
             }

   \date{Received 3 November 2009 / Accepted 11 January 2010}

 \abstract{X-ray spectra of active galactic nuclei (AGN) are usually described as power law spectra, characterized by the spectral slope $\alpha$ or photon index $\Gamma$.
Here we discuss the X-ray spectral properties within the framework of clumpy accretion flows, and estimate the power law slope as a function of the source parameters.
We expect harder spectra in massive objects than in less massive sources, and steeper spectra in higher accretion rate systems.
The predicted values of the photon index  cover the range of spectral slopes typically observed in Seyfert galaxies and quasars.
The overall trends are consistent with observations, and may account for the positive correlation of the photon index with Eddington ratio (and the possible anticorrelation with black hole mass) observed in different AGN samples.
Spectral properties are also closely related to variability properties.
We obtain that shorter characteristic time scales are associated with steeper spectra. This agrees with the observed `spectral-timing' correlation.}
     

   \authorrunning{ W. Ishibashi \and T. J.-L. Courvoisier}
   \titlerunning{X-ray power law spectra in AGN}

   \maketitle
%

\section{Introduction}

Hard X-ray spectra of active galactic nuclei (AGN) are usually decomposed into an underlying primary power law, a Compton reflection hump and an iron line. 
The power law component, of the form $I_{f} \propto f^{-\alpha}$, is characterized by the spectral slope $\alpha$ or photon index $\Gamma = \alpha + 1$. 
Power law photon indices have been measured in different AGN samples, with typical values lying in the $\Gamma \sim 1.5-2.5$ range \citep{N_P_1994, R_T_2000, P_et_2005, PK_et_2005}.
 
The relationship between power law slope and source parameters like black hole mass and accretion rate, has been extensively analyzed for different AGN samples. 
The photon index is found to be anticorrelated with black hole mass in samples drawn from the Palomar Green (PG) survey \citep{P_et_2004, P_et_2005}, but this anticorrelation seems to disappear in higher luminosity objects \citep{S_et_2006}. 
A positive correlation between photon index and Eddington ratio $L/L_{E}$ was reported in several papers \citep{P_et_2004, P_et_2005, S_et_2006, S_et_2008}. 
But non-monotonic trends of the photon index on source parameters were also suggested \citep{K_et_2008}.
A difference in the power law slope was observed in different AGN classes, in particular Seyfert 2 galaxies are found to have harder spectra than Seyfert 1
galaxies \citep{Z_et_1995, Z_et_2000}.
Furthermore, \citet{Z_et_1999} reported a significant correlation between reflection component and power law slope in Seyfert galaxies. 
More recently, a `spectral-timing' correlation was discovered in AGNs \citep{P_et_2009}, suggesting a close connection between X-ray spectral and variability properties.  

The physical process at the origin of X-ray emission is assumed to be Comptonization of seed photons by energetic electrons in a hot plasma cloud. The observed X-ray spectra are indeed well represented by Comptonization models \citep[and references therein]{T_1994}. 
\citet{T_L_1995} showed that power law spectra are the exact solution of the radiative kinetic equation and give an analytical solution for the spectral slope, valid without restriction on plasma physical parameters. 

Here, we study the X-ray spectral properties within the framework of our cascades of shocks model, previously discussed \citep[][hereafter Paper I]{C_T_2005, Paper_I}.
In this picture, radiation is emitted as a result of shocks between elements (clumps) within the inhomogeneous accretion flow.
Optically thick shocks give rise to optical/UV photons, which are later upscattered by hot electrons in optically thin shocks. These second generation shocks are responsible for the X-ray emission. 
Here we compute the X-ray power law index $\Gamma$ as a function of the object parameters in our model and compare it with observations. 
In particular, we study the relationship between X-ray spectral slope and source physical parameters, black hole mass and accretion rate. 
We also discuss X-ray spectral properties in connection with the X-ray variability properties analyzed in a previous paper \citep[hereafter Paper II]{Paper_II}.   

The present paper is organized as follows. We recall the main relevant features of our cascades of shocks model in Sect. 2. 
In Sect. 3 we compute analytical estimates of the power law index.
A comparison of our results with observations is given in Sect. 4.  
We analyze the relationship between X-ray spectral and variability properties in Sect. 5. 
We discuss resulting implications and other more speculative aspects in Sect. 6. 
The main results are summarized in the Conclusion.


\section{Cascades of shocks in clumpy accretion flows}

We consider the accretion flow onto the central black hole to be very inhomogeneous and model it as formed by individual clumps of matter interacting with one another. 
At a typical distance of $\sim$100 Schwarzschild radii, clumps of mass $M_{c} \cong M_{33} \times 10^{33}$g move at the local free-fall velocity $v_{ff} \cong 3 \times 10^{9} \zeta_{UV}^{-1/2}$cm/s (where $\zeta_{UV} = \zeta/100$, and the radial distance is expressed in units of the Schwarzschild radius, $R = \zeta R_{S}$). The interaction between two such clumps and the resulting shock leads to a gas cloud in rapid expansion, similar to a supernova explosion. 
The shock occurs in optically thick conditions and a fraction $\eta_{rad} = \eta_{1/3} \times 1/3$ of the kinetic energy of the colliding clumps is converted into radiative energy. As a result, optical/UV emission is radiated at the photosphere of the expanding cloud. The photospheric radius is given by
\begin{equation}
R_{max} \cong 3 \times 10^{15} M_{33}^{1/2} \zeta_{UV}^{-1/4} \, \mathrm{cm} \, .
\end{equation}

In the inner regions of the object, expanding clouds overlap with one another, leading this time to optically thin shocks. 
In these shocks, electrons gain energy through Coulomb collisions with hot protons and cool through Compton emission. 
Seed photons for Compton cooling are provided by optical/UV photons emitted in the optically thick event.
The average electron energy is set by the balance between Coulomb heating and Compton cooling rates:
\begin{equation}
L_{Coulomb} = L_{Compton} \, . 
\end{equation}
The resulting electron temperature is of a few hundred keV, and thus Compton cooling of the hot electrons gives rise to X-ray emission. 

In Paper I, we distinguished two classes of objects, Class S and Class Q, depending on the volume filling factor of the expanding clouds with respect to the local region size. The volume filling factor $\epsilon$ was defined as
\begin{equation}
\epsilon =   \left( \frac{R_{max}}{100 \zeta_{UV} R_{S}} \right)^{3} \cong 10^{-3} M_{33}^{3/2} \zeta_{UV}^{-15/4} 
\left( \frac{M_{BH}}{10^{9} M_{\odot}} \right)^{-3} \, .
\end{equation}

Class Q objects have large black hole mass ($M_{BH} \gtrsim 10^{9} M_{\odot}$) for which the volume filling factor is small ($\epsilon \ll 1$), while Class S objects are characterized by a large filling factor ($\epsilon \sim 1$), associated with smaller black hole masses ($M_{BH} \lesssim 10^{8} M_{\odot}$). 
We identified Class Q objects with massive and luminous quasars, and Class S sources with less luminous Seyfert galaxies. 
In Class S, the black hole mass is expressed in units of $M_{BH} = M_{8} \times 10^{8} M_{\odot}$, and the accretion rate in units of 
$\dot{M} = \dot{M}_{0} \times 1 M_{\odot}/\textrm{yr}$; in Class Q, the black hole mass is expressed in units of $M_{BH} = M_{9} \times 10^{9} M_{\odot}$, and the accretion rate in units of 
$\dot{M} = \dot{M}_{1} \times 10 M_{\odot}/\textrm{yr}$. 

We further defined two subcases, Cases A and B, comparing the electron accretion and radiation time scales. The separation depends on whether the ratio of the Compton cooling time ($t_{cool}$) over the local dynamical time ($t_{dyn}$) is smaller or greater than unity.
In Class S, this condition can be written as (see Eq. (22) in Paper I)
\begin{equation}
\frac{t_{cool}}{t_{dyn}} < 1 \Leftrightarrow \left( \frac{\dot{M}}{1 M_{\odot}/yr} \right) / \left( \frac{M_{BH}}{10^{8} M_{\odot}} \right) 
> 0.4 \, \eta_{1/3}^{-1} \zeta_{UV}^{3/2} \, .
\end{equation}
We see that the time scale condition translates into a condition on the ratio of accretion rate to central mass, $\dot{M}/M_{BH}$.
This parameter can be related to the Eddington ratio $L/L_{E}$, since
\begin{equation}
\frac{L}{L_{E}} \propto \frac{\dot{M}}{M_{BH}} \, ,
\end{equation}
where $L_{E}$ is the Eddington luminosity.

Case A is defined by the condition $t_{cool} > t_{dyn}$, and Case B by the opposite condition, $t_{cool} < t_{dyn}$.
In other words, for a given black hole mass, there are two cases divided into lower accretion rate (Case A) and higher accretion rate (Case B) systems. 
As shown in Paper I, the time scale condition implies that only Case A is appropriate for Class Q objects.

In Paper II, we introduced a characteristic X-ray time scale and calculated that its dependence on black hole mass and accretion rate is of the form
\begin{equation}
t_{X} \propto \frac{M_{BH}^{2}}{\dot{M}} \, .
\end{equation}
The model time scale is thus shorter for small black hole mass and/or high accretion rate. 
We have shown that this model dependence on black hole mass and accretion rate is in excellent agreement with the empirical relation obtained by \citet{McH_et_2006}.


\section{Estimation of the power law spectral slope}

The spectral shape of the X-ray emission, and in particular the value of the spectral slope, is determined by two plasma physical parameters: the electron temperature and the medium optical depth. 
The dimensionless electron temperature is defined as
\begin{equation}
\Theta = \frac{kT_{e}}{m_{e}c^{2}} \, , 
\end{equation}
and is given by the balance between Coulomb heating and Compton cooling rates.
The optical depth of the cloud is given by
\begin{equation}
\tau = n_{e} \sigma_{T} R \, ,
\label{tau}
\end{equation}
where $n_{e}$ is the electron number density, $\sigma_{T}$ the Thomson cross section, and $R$ the typical size of the upscattering region. 

We compute the power law spectral slope using estimates of the electron temperature and optical depth obtained within our model. 
We follow \citet{T_L_1995} who give the exact analytical solution for the spectral index without restriction on plasma temperature and optical depth. 
In the non-relativistic limit the spectral slope is given by
\begin{equation}
\alpha = \sqrt{9/4 + \beta/\Theta} - 3/2 \, , 
\label{NR_alpha}
\end{equation}
where the parameter $\beta$ is only a function of the optical depth and the geometry (plane or spherical). 
The spectral index in the relativistic case is given by
\begin{equation}
\alpha = \frac{\beta -\ln d_{0}\left( \alpha \right)}{\ln \left( 4 \Theta^{2} \right)} \, ,
\label{R_alpha}
\end{equation}
where the coefficient $d_{0}\left( \alpha \right)$ is computed through a transcendental equation.

\subsection{Class S}

The electron temperature in Class S was calculated in Paper I and is on the order of  
\begin{equation}
\Theta = \frac{kT_{e}}{m_{e}c^{2}} \sim 0.6 \, .
\label{T_e}
\end{equation}
We also estimated the typical electron density, which we parametrize here as $n_{e} = n_{S} \cdot 1.2 \times 10^{8} \mathrm{cm^{-3}}$. 

In Case A, the cooling time scale is longer than the dynamical time scale, and the size of the upscattering region is therefore given by the size of the shock region, $R \cong 3 \times 10^{15} \zeta_{UV} M_{8} \mathrm{cm}$. 
In Case B, the cooling time scale is shorter than the dynamical time scale, the size of the Comptonization region is then given by the distance over which electrons travel at the local free-fall speed during the cooling time, $R = v_{ff} \cdot t_{cool} \cong 1.2 \times 10^{15} \eta_{1/3}^{-1} \zeta_{UV}^{5/2} \dot{M}_{0}^{-1} M_{8}^{2} \mathrm{cm}$. 

The optical depths for the two cases are obtained with Eq. (\ref{tau})
\begin{equation}
\textrm{Case A:} \quad \tau \cong 0.2 \, n_{S} \zeta_{UV} M_{8}
\label{tau_A}
\end{equation}
\begin{equation}
\textrm{Case B:} \quad \tau \cong 0.1 \, n_{S} \eta_{1/3}^{-1} \zeta_{UV}^{5/2} \dot{M}_{0}^{-1} M_{8}^{2} \, . 
\label{tau_B}
\end{equation}
We can now use these estimates of the optical depth to compute the parameter $\beta$ in the spherical geometry, following \citet{T_L_1995}. 
Inserting the value of the electron temperature (eq. \ref{T_e}) and the respective values of the $\beta$ parameter in eq. (\ref{NR_alpha}), we obtain the spectral slopes in the two cases
\begin{equation}
\textrm{Case A:} \quad \alpha \sim 0.9 \Longrightarrow \Gamma \sim 1.9
\end{equation}
\begin{equation}
\textrm{Case B:} \quad \alpha \sim 1.1 \Longrightarrow \Gamma \sim 2.1 \, .
\end{equation}
We see that the power law spectrum is steeper in Case B ($\Gamma \sim 2.1$) than in  Case A ($\Gamma \sim 1.9$). 
When computing the spectral slope as a function of the optical depth, we note for a given electron temperature that the power law spectrum becomes steeper for small optical depths. Since the optical depth is smaller in Case B than in Case A, Case B objects are predicted to have steeper spectra than Case A objects.

We recall that Case A and Case B are distinguished by the time scale condition, expressed in terms of the $\dot{M}/M_{BH}$ parameter, which may be considered as a measure of the Eddington ratio $L/L_{E}$.
We thus have two cases: the lower $\dot{M}/M_{BH}$ case (Case A) and the higher $\dot{M}/M_{BH}$ case (Case B). 
Steeper spectra are predicted in the latter case, where the optical depth decreases with increasing accretion rate for a given black hole mass. 
Thus an object accreting at the Eddington limit can have a very steep spectrum, with a photon index around $\Gamma \sim 2.3$.

\subsection{Class Q}

The electron temperature in Class Q was also estimated in Paper I and is on the order of
\begin{equation}
\Theta = \frac{kT_{e}}{m_{e}c^{2}} \sim 1.4 \, . 
\end{equation}
The upscattering region size is roughly given by $R \cong 3 \cdot 10^{15} \zeta_{X} M_{9} \mathrm{cm}$ (where $\zeta_{X} = \zeta / 10$), the optical depth is
estimated as in Class S
\begin{equation}
\tau \cong 0.4 \, n_{Q} \zeta_{X} M_{9} \, ,
\label{tau_Q}
\end{equation}
with an appropriate parametrization of the electron density $n_{e} = n_{Q} \cdot 1.9 \times 10^{8} \mathrm{cm^{-3}}$.
Here we estimate the spectral slope in the relativistic regime (Eq. \ref{R_alpha}).
Starting from the value of $\alpha$ obtained in the non-relativistic limit and after a few iterations, the spectral index converges toward a value of $\alpha \sim 0.5$, leading to a photon index of $\Gamma \sim 1.5$. 
We observe that the photon index is smaller in Class Q objects than in Class S sources.
This implies that the power law spectrum is harder in massive objects compared to less massive sources. 
Observational measurements of black hole masses are believed to be accurate to within a factor of a few. 
Varying the mass parameter in Eq.(\ref{tau_Q}) leads to some variations of the photon index, nevertheless the resulting spectra are always harder than spectra in Class S sources. 
Moreover, the trend indicating a hardening of the spectrum with increasing central mass is preserved within Class Q objects.


\section{Comparison with observations}

\subsection{Power law index range}

X-ray power law spectra have been studied for a long time, and photon indices in the hard energy band have been measured in different AGN samples. 
\citet{N_P_1994} found a mean X-ray photon index in the range $\Gamma \cong 1.9-2.0$ for a sample of 27 Seyfert galaxies, based on Ginga observations. 
\citet{R_T_2000} analyzed a large sample of 62 quasars (35 radio-loud and 27 radio-quiet) and observe that the majority of quasars have a photon index in the range $\Gamma \sim 1.5-2.1$.  
Similar values are found for a sample of 40 quasars (35 radio-quiet and 5 radio-loud), observed with XMM-Newton, from the Palomar Green Bright Quasar Survey (PG BQS), with an average hard X-ray power law index of $\Gamma \cong 1.89\pm0.11$ for the radio-quiet sub-sample (Piconcelli et al. 2005).  
Thus observations indicate that the X-ray power law index spans a limited range, roughly between 1.5 and 2.5 \citep{S_et_2006, S_et_2008}. 
Combining Class S and Class Q objects, we obtain power law indices in the range $\Gamma \sim 1.5-2.1$, covering the typically observed values.

\subsection{Dependence of the photon index on source parameters}

The relationship between X-ray spectral slope, black hole mass and Eddington ratio has been studied by many authors. The hard X-ray photon index is observed to be correlated with the Eddington ratio and to anticorrelate with the black hole mass in a sample of 21 low-redshift Palomar Green quasars \citep{P_et_2004}. The trend indicating a steepening of the observed spectrum for lower black hole mass and higher accretion rates is confirmed for a larger sample of PG quasars \citep{P_et_2005}. 
\citet{S_et_2006} analyzed a sample of 30 moderate-to-high luminosity radio-quiet AGNs and reported a significant correlation between the hard X-ray photon index and normalized accretion rate ($L/L_{E}$), while they observed that the $\Gamma-M_{BH}$ anticorrelation disappears when including the high-luminosity objects. Thus they argued that the hard X-ray photon index depends primarily on the accretion rate. This conclusion was reinforced by \citet{S_et_2008}, who suggested that the photon index $\Gamma$ may be used as an accretion rate indicator. 
Similar results, i.e. a $\Gamma-L/L_{E}$ correlation and a weaker $\Gamma-M_{BH}$ anticorrelation, have been obtained in larger samples from the Sloan Digital Sky Survey (SDSS), but with some caveats related to emission lines used for the black hole mass estimates \citep{K_et_2008, R_et_2009}. 
Summarizing, observations tend to agree on the strong correlation between hard X-ray photon index and some form of accretion rate, while a weaker anticorrelation between $\Gamma$ and black hole mass cannot be ruled out.  

In our picture, smaller optical depths and hence steeper spectra are expected in higher accretion rate systems (Case B) compared with lower accretion rate objects (Case A). This agrees with the observed $\Gamma-L/L_{E}$ correlation. 
Furthermore, very steep spectra ($\Gamma \sim 2.3$) are expected for low-mass objects accreting at high accretion rates (Case B). 
Extremely steep spectra are in fact observed in the particular class of Narrow-line Seyfert 1 (NLS 1) galaxies, which are believed to be powered by small black holes accreting at rates close to the Eddington limit. 
Therefore extreme Case B objects might be associated with NLS 1 galaxies.


\section{Relationship between X-ray spectral and variability properties}

\subsection{Characteristic variability time scale}

X-ray spectral properties can be linked with X-ray variability properties. 
The temporal structure is often analyzed in terms of the power spectral density (PSD). 
The power spectrum is modeled by a power law of the form $P_{\nu} \propto \nu^{\alpha_{T}}$, where $\nu = 1/T$ is the temporal frequency. 
A break is observed in the power spectrum at a given break frequency, $\nu_{B}$. The associated time scale, $T_{B} = 1/\nu_{B}$, represents a characteristic X-ray time scale. \\

It is well known that narrow-line Seyfert 1 galaxies (NLS 1) have higher break frequencies, or equivalently shorter characteristic time scales, than broad-line galaxies (BLS 1) of the same mass. 
It has been argued that for the same black hole mass, the break frequency should be $\sim$20 times higher in NLS 1 galaxies than in BLS 1 objects \citep{P_2004}, suggesting that the break time scale might depend on additional parameters.
This led \citet{McH_et_2006} to propose a scaling relationship in which the characteristic time scale has an explicit dependence on both black hole mass and accretion rate. 
The observational relationship is of the form
\begin{equation}
T_{B} \approx M_{BH}^{1.12}/\dot{m}_{E}^{0.98} \, ,
\end{equation}
where $\dot{m}_{E} = L_{bol}/L_{E}$, with $L_{bol}$ the bolometric luminosity and $L_{E}$ the Eddington luminosity. \\

\citet{N_et_2009} raised the possibility that the dependence of the variability properties on the accretion rate might not be continuous. They instead suggest a bimodal distribution, with BLS 1 and NLS 1 belonging to two distinct populations (with a scaling factor differing by a factor of $\sim$20).  
This bimodal behavior is attributed to a bimodal dependence on the soft X-ray spectral slope rather than to a dependence on accretion rate.

In our picture, NLS 1 galaxies could be associated with the highest accretion rate Case B objects, while BLS 1 objects having lower accretion rates might be linked with Case A sources.
But the transition between the two cases will probably not result in a bimodal distribution in the variability properties.
Moreover, the distinction between the two cases is related to the accretion rate (through the $\dot{M}/M_{BH}$ parameter), differing from the interpretation suggested by \citet{N_et_2009}. 
 
\subsection{`Spectral-timing' relation}

Recently, \citet{P_et_2009} report for the first time a `spectral-timing' correlation in AGNs: the average X-ray spectral slope is positively correlated with the normalized characteristic frequency, defined as $\nu_{norm} = \nu_{B} \times M_{BH}$. 
This correlation implies that objects which on average have steeper spectra also have higher characteristic frequencies, or equivalently shorter characteristic time scales.
Using the result of \citet{McH_et_2006}, the `spectral-timing' relation suggests that for a given black hole mass, objects with higher accretion rate relative to Eddington also have steeper spectra.
\citet{P_et_2009} argue that the accretion rate determines both the PSD break frequency and the X-ray spectral slope.

In Paper II, we studied the X-ray variability properties within our model, focusing on Class S objects since timing analyses are mainly performed on local Seyfert 1 galaxies. We derived a characteristic X-ray time scale of the form $t_{X} \cong 5 \, \eta_{1/3}^{-1} \zeta_{UV}^{3} \dot{M}_{0}^{-1} M_{8}^{2}$d. 
The model time scale is shorter for small black hole mass and/or high accretion rate. 
The predicted dependence on black hole mass and accretion rate of the form $t_{X} \propto M_{BH}^{2} / \dot{M}$ precisely reproduces the empirical relation $T_{B} \propto M_{BH}^{2.1}/\dot{M}^{0.98}$ of \citet{McH_et_2006}. \\

In Case B, the optical depth decreases with increasing accretion rate (Eq. \ref{tau_B}), and we obtain steeper spectra for high accretion rate objects, whereas in Case A we do not have an explicit dependence on the accretion rate parameter (Eq. \ref{tau_A}). 
However, we recall that Case A is formed by lower accretion rate systems (Case A is indeed defined by the condition $\left( \frac{\dot{M}}{1 M_{\odot}/yr} \right) / \left( \frac{M_{BH}}{10^{8} M_{\odot}} \right) < 0.4 \, \eta_{1/3}^{-1} \zeta_{UV}^{3/2}$), which tend to have longer characteristic time scales.
In  Case B, the characteristic time scale and the optical depth are related through their dependence on black hole mass and accretion rate. 
With Eq. (\ref{NR_alpha}), we can thus plot the spectral slope $\alpha$ or equivalently the photon index $\Gamma$ as a function of the characteristic time scale for a given electron temperature (Fig. 1).
We note that Case B objects with steeper spectra are associated with shorter characteristic time scales and vice versa.
Case A objects would be confined towards higher values of $t_{X}$ with a scatter around a characteristic $\Gamma$ value. 

Combining the above results, we obtain that the characteristic time scale is shorter and the spectrum steeper in higher accretion rate systems.
This agrees with the observed `spectral-timing' relation. 
The shortest characteristic time scale along with the steepest spectrum should be obtained in the case of a low-mass object accreting at a high rate.
Such extreme properties are indeed observed in NLS 1 galaxies. Thus black hole mass and accretion rate seem to determine both the X-ray variability and spectral properties. 

   \begin{figure}
   \centering
   \includegraphics[width=0.5\textwidth]{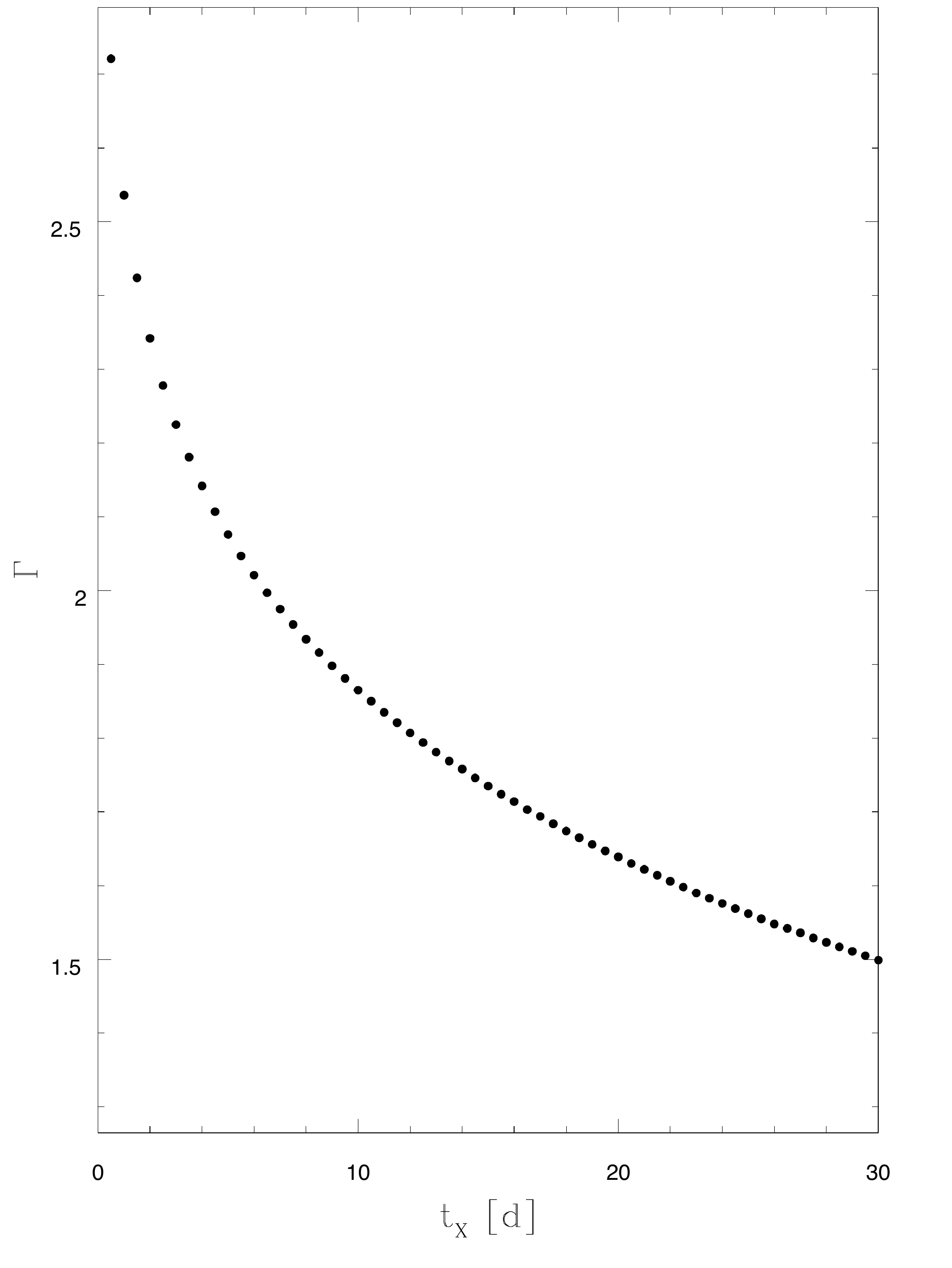}
      \caption{Power law photon index $\Gamma$ as a function of the characteristic variability time scale $t_{X}$ for Class S, Case B objects. 
      Short characteristic time scales are associated with steep spectra, while longer time scales are associated with harder spectra.  
 }      
         \label{figure}
   \end{figure}


\section{Discussion}

\subsection{The $\Gamma-L/L_{E}$ relation and the unification scenario}

The correlation between photon index $\Gamma$ and accretion rate, expressed in terms of the Eddington ratio $L/L_{E}$, has been confirmed in numerous studies. 
The usual interpretation of the $\Gamma-L/L_{E}$ correlation is that an enhancement in the accretion rate leads to an increase in the seed photon emission, which in turn leads to efficient cooling and steeper spectra.
In our picture, the cooling time scale is shorter than the dynamical time scale in the higher accretion rate regime (Case B), and thus Compton cooling of the hot electrons is very efficient.

According to the unified model of AGNs, the distinction between type 1 and type 2 objects is given by the orientation of the obscuring torus with respect to the observer \citep{A_1993}. In this scenario, high-energy spectra should be similar in both type of objects since the absorption by the molecular torus should not affect high energy spectra above 20-30 keV, unless the source is Compton-thick. 
However, a difference in high-energy spectra has been observed between Seyfert type 1 and type 2 objects, with Seyfert 2 galaxies having harder spectra than Seyfert 1 galaxies \citep{Z_et_1995, Z_et_2000}. 
\citet{D_C_2003} analyzed hard X-ray spectra of Seyfert galaxies observed with BeppoSAX/PDS and found differences between type 1 and type 2 sources that cannot only be explained with orientation effects.
More recently, \citet{M_et_2008} report softer spectra in Seyfert 1 galaxies ($\Gamma \sim 2.1$) compared with Seyfert 2 galaxies ($\Gamma \sim 1.9$). In addition, they found a difference in the mean Eddington ratio, with type 1 objects having higher values than type 2 sources. 
Thus the authors argue that the steeper spectra observed in Seyfert 1 galaxies may be attributed to an intrinsic softening with increasing accretion rate, consistent with the overall $\Gamma-L/L_{E}$ correlation.
This difference in Eddington ratios between type 1 and type 2 objects has been also confirmed in a larger sample of 71 Seyfert galaxies observed with INTEGRAL \citep{B_et_2009}. 
The difference between type 1 and type 2 objects would then be given by the difference in their mean accretion rate, and the two Seyfert sub-classes would belong to two different accretion rate regimes. 
If this interpretation is correct, we might identify different model cases with different observed categories, based on the accretion rate parameter.
The lower accretion rate Seyfert 2 objects could then be associated with Class S, Case A objects, while the higher accretion rate Seyfert 1 galaxies might be linked with Class S, Case B objects. 
But other authors claim the difference in spectral slope between type 1 and type 2 objects is not significant and compatible within uncertainties, when considering more complex model fitting \citep{B_et_2009}. 
It then seems that the exact value of the spectral slope depends on the complexities of the model used to fit the data, and the issue remains still open.

\subsection{The $\Gamma-M_{BH}$ relation and the radio-loudness dichotomy}

While the $\Gamma-L/L_{E}$ correlation seems secure, the dependence of the photon index on black hole mass is less clear. 
A number of studies report an anticorrelation between $\Gamma$ and $M_{BH}$ \citep{P_et_2004, P_et_2005}, but this anticorrelation has been claimed to disappear when including high-luminosity objects \citep{S_et_2006}. 
The $\Gamma-M_{BH}$ anticorrelation is confirmed in a large sample of 153 radio-quiet AGNs when using the $H\beta$ line as a black hole mass estimator, but an opposite trend is found when using the $CIV$ line \citep{K_et_2008}. Indeed the use of some emission lines as black hole mass estimators has been questioned, in particular it has been argued that the $CIV$ line is not a reliable mass estimator \citep{R_et_2009}. 
Thus the $\Gamma-M_{BH}$ anticorrelation cannot be ruled out and seems still significant, although clearly weaker than the $\Gamma-L/L_{E}$ relation. 
Here we obtain harder spectra in Class Q ($\Gamma \sim 1.5$) than in Class S ($\Gamma \sim 1.9-2.1$) objects.
Thus if the $\Gamma-M_{BH}$ anticorrelation is real, the predicted hardening of the spectrum with increasing black hole mass is consistent with observations.

A systematic difference in the X-ray spectral slope, related to the radio-loudness of the source, has been known for a long time. 
Radio-loud objects have harder spectra ($\Gamma \sim 1.6$) than radio-quiet sources ($\Gamma \sim 1.9$) \citep{R_T_2000, PK_et_2005, P_et_2005}.
The flatter spectra observed in radio-loud AGNs are usually attributed to an enhanced X-ray emission from the jet component. 
Note however that variability analysis in 3C 273 has shown no correlation between X-ray emission and radio-mm emission from the jet \citep{So_et_2008}. 
On the other hand, a connection between radio-loudness and black hole mass has also been observed.  Analyzing a sample of PG quasars, \citet{L_2000} noted that nearly all objects with $M_{BH} > 10^{9} M_{\odot}$ are radio-loud, while sources with $M_{BH} < 3 \cdot 10^{8} M_{\odot}$ are radio-quiet. He thus suggested a radio-loudness bimodality related to black hole mass.
This implies that the difference in $\Gamma$ between radio-loud and radio-quiet objects may also be related to a difference in their respective black hole mass. 
The association of radio-loud objects with massive black holes is confirmed in larger samples from the Sloan Digital Sky Survey \citep{McL_J_2004}. 
The above results imply that X-ray spectra should be harder in radio-loud, hence more massive objects, a trend in agreement with the $\Gamma-M_{BH}$ anticorrelation. 
The fact that radio-loudness seems to be restricted to objects with central masses larger than a critical value points toward a still unknown relation between jet activity and black hole mass.
However, it is clear that other parameters like accretion rate and black hole spin are required to account for the jet activity.


\subsection{The reflection component}

A Compton reflection component creating a hump above the underlying power law, with a peak around 30 keV is often observed in the X-ray spectra of AGNs. This spectral component is usually interpreted as due to reflection of hard X-ray photons off a cold optically thick medium, such as the accretion disk and/or the obscuring torus. 

\citet{R_T_2000} observe that the amount of reflection is much weaker in luminous, high-redshift objects compared with lower-luminosity Seyfert galaxies. Indeed no Compton reflection hump has been detected in a sample of 29 high-redshift quasars observed with XMM-Newton \citep{PK_et_2005}. Similarly, \citet{S_et_2008} found no significant reflection component in most of their high-redshift, high-luminosity sources. 
It seems that the importance of the Compton reflection component declines with increasing luminosity, i.e. it is weaker in high-luminosity quasars than in low-luminosity Seyfert galaxies. 
This trend cannot be uniquely attributed to Doppler boosting of the jet component, since the effect is observed in both radio-loud and radio-quiet sources. 

In our picture, the reflection component can be discussed in terms of the volume filling factor of the expanding clumps. 
We predict a weak reflection component in Class Q objects, in which the volume filling factor of the optically thick clouds is very small ($\epsilon \ll 1$), while a stronger reflection component is expected in the less massive Class S objects, where the covering fraction is close to unity ($\epsilon \sim 1$). 
This behavior is qualitatively consistent with the observed decline in importance of the reflection component with increasing luminosity of the source. 

In addition, the importance of the reflection component seems to be related to the X-ray spectral shape: a significant correlation between power law slope and amount of reflection has been observed in Seyfert galaxies \citep{Z_et_1999}. 
Softer spectra are associated with a strong reflection component, while less reflection is observed in objects with harder spectra. 
In general, Seyfert galaxies are characterized by quite soft spectra together with large reflection.
Such properties are expected in Class S objects, which we have identified with Seyfert galaxies. 
If the correlation between spectral slope and amount of reflection holds for higher-luminosity objects, then Class Q objects might be associated with sources having hard spectra and weak reflection.


\section{Conclusion}

We have studied the X-ray spectral properties within the clumpy accretion flow scenario. 
In this picture, X-rays are directly emitted as a consequence of the Compton cooling process of hot electrons in the optically thin shocks. 
We compute the power law photon index using estimates of the electron temperature 
and optical depth derived within our model and obtain values of $\Gamma$ corresponding to the typically measured values.
The predicted trends of the photon index with black hole mass and accretion rate may reproduce the observed $\Gamma-L/L_{E}$ and $\Gamma-M_{BH}$ relations.
Spectral and variability properties are closely related, with steeper spectra being associated with shorter characteristic time scales. 
In our picture, such `spectral-timing' relation is derived in a natural way, through the dependence on source parameters. 
Thus both X-ray spectral and variability characteristics seem to be mainly determined by two parameters, black hole mass and accretion rate.


\bibliographystyle{aa}
\bibliography{biblio}

\end{document}